\documentclass{jnmp}
 
\JNMPnumberwithin{equation}{section} 
 
\usepackage{amsmath}

\begin{document} 
 
\renewcommand{\evenhead}{D~B~Fairlie} 
\renewcommand{\oddhead}{A Universal Solution} 
 
\thispagestyle{empty} 
 
\FirstPageHead{9}{3}{2002}{\pageref{fairlie-firstpage}--\pageref{fairlie-lastpage}}{Letter} 
 
\copyrightnote{2002}{D~B~Fairlie} 
 
\Name{A Universal Solution}\label{fairlie-firstpage} 
 
\Author{D~B~FAIRLIE} 
 
\Address{Department of Mathematical Sciences, University of Durham, Durham DH1 3LE\\ 
E-mail: david.fairlie@durham.ac.uk} 
 
\Date{Received January 7, 2002; Revised(1) March 26, 2002; 
Revised(2) April 2, 2002;\\ Accepted April 4, 2002}

\begin{abstract} 
\noindent 
The phenomenon of an implicit function which solves a large 
set of second order partial differential 
equations obtainable from a variational principle is explicated by the introduction of a class of 
universal solutions to the equations derivable from an arbitrary Lagrangian 
 which is homogeneous of 
weight one in the field derivatives. This result is extended to many fields. The imposition of 
Lorentz invariance makes such Lagrangians unique, and equivalent 
 to the Companion Lagrangians introduced 
in~\cite{baker}. 
\end{abstract} 
 
\section{Introduction} 
The simplest example of a universal solution is that of a linear function, i.e. 
\[ 
f(x_i)=a_1x_1+a_2x_2+\cdots +a_nx_n 
\] 
satisfies all partial differential equations in $n$ variables each 
of whose terms contains at least one factor which is a derivative 
of second order or higher. The purpose of this article is to 
display a large class of functions of which the linear function is 
a particular case which are solutions of a set, infinite in 
general, of partial differential equations derived by variation of 
a Lagrangian. The result is given by the following theorem, and is 
then extended to the case of several unknowns. It is shown that 
the functions also provide a~solution to the iterated variations 
of this class of Lagrangians, which terminates in the so-called 
Universal Field Equation. Finally, the first order formulation of 
the equations of motion provides further understanding of the 
mechanism behind the result claimed. 
 
\medskip 
 
\noindent 
{\bf Theorem.} {\it 
Suppose $\phi(x_i)$, $i=1,\dots,n$ is a differentiable function of $n$ variables 
$x_i$. 
 Let~$\phi_j$ denote $\frac{\partial\phi}{\partial x_j}$, $j=1,\dots,n$ 
and let $F^j(\phi)$ be any arbitrary 
differentiable functions of the single argument $\phi$ subject to the single constraint 
\begin{equation} 
\sum_{j=1}^{j=n}x_jF^j(\phi)={\rm constant}.\label{one} 
\end{equation} 
Then an implicit solution of this constraint for $\phi$ 
is a solution to any equation of motion for $\phi$ derivable from a Lagrangian of the form 
\[{\cal L}= {\cal L}(\phi,\phi_j),\] 
where ${\cal L}$ is homogeneous of weight one in the first 
derivatives of $\phi$, i.e.\ 
\begin{equation} 
\sum_{j=1}^{j=n}\phi_j\frac{\partial{\cal L}}{\partial \phi_j}={\cal L}.\label{two} 
\end{equation}} 
 
The proof of this result is relatively easy. 
First of all from (\ref{one}) it follows by differentiation that 
\begin{gather} 
\frac{\partial\phi}{\partial x_j} = -\frac{F^j}{\sum x_i (F^i)'},\nonumber\\ 
\frac{\partial^2\phi}{\partial x_j\partial x_k}=\frac{F^j(F^k)'+F^k(F^j)'} 
{\left(\sum x_i(F^i)'\right)^2}+ 
\frac{F^jF^k\left(\sum x_r(F^r)''\right)}{\left(\sum x_i(F^i)'\right)^3}  \nonumber\\ 
                               \phantom{\frac{\partial^2\phi}{\partial x_j\partial x_k}} = -\frac{\phi_j(F^k)'+\phi_k(F^j)'} 
{\sum x_i(F^i)'}+ 
\frac{\phi_j\phi_k\left(\sum x_r(F^r)''\right)}{\sum x_i(F^i)'}.\label{result} 
\end{gather} 
Here a prime denotes differentiation once with respect to the argument $\phi$, 
i.e.\ $(F^j)'=\frac{dF^j}{d\phi}$. 
The equation of motion for ${\cal L}$ is 
\begin{gather} 
\frac{\partial{\cal L}}{\partial \phi}-\frac{\partial}{\partial x_j} 
\frac{\partial{\cal L}}{\partial \phi_j}\nonumber\\ 
\qquad {}=\frac{\partial{\cal L}}{\partial \phi}-\frac{\partial \phi}{\partial x_j}\frac{\partial^2{\cal L}} 
{\partial \phi_j\partial\phi} -\frac{\partial^2\phi}{\partial x_j\partial x_k}\frac{\partial^2{\cal L}} 
{\partial \phi_j\partial\phi_k}=0. 
\end{gather} 
The first two terms cancel because 
$\frac{\partial{\cal L}}{\partial \phi}$ is also homogeneous of 
degree one in $\phi_j$ leaving 
\begin{equation}\frac{\partial^2\phi}{\partial x_j\partial x_k}\frac{\partial^2{\cal L}} 
{\partial \phi_j\partial\phi_k}=0\label{eqnmot} 
\end{equation} 
as equation of motion. 
But from differentiating (\ref{two}) 
\[ 
\sum_j\frac{\partial\phi}{\partial x_j}\frac{\partial^2{\cal L}} 
{\partial \phi_j\partial\phi_k}=0. 
\] 
Using this result, together with (\ref{result}) which expresses 
the second derivatives of $\phi$ in terms of the first, the  
theorem is established. Note that a characteristic feature of this 
equation of motion is that it is covariant, i.e.\ if~$\phi$ is a 
solution so is any function of $\phi$ and this feature is manifest 
in part of the arbitrariness of the universal solution. 
 
\section{Multifield extension} 
The theorem extends to 
 the  the case where the Lagrangian depends upon the first derivatives of 
several fields $\phi^\alpha$, and satisfies the following orthogonality 
relations for the gradients of each field 
\begin{equation} 
\sum_j\frac{\partial\phi^\alpha}{\partial x_j} 
\frac{\partial {\cal L}}{\partial \frac{\partial\phi^\beta}{\partial x_j}}=\delta_{\alpha\beta} 
{\cal L}.\label{hom} 
\end{equation} 
(These are somewhat stronger conditions than to demand homogeneity 
in the first derivatives of each $\phi^\alpha$.) 
In this situation the equations of motion are 
\begin{equation} 
\sum_\beta\frac{\partial^2\phi^\beta}{\partial x_j\partial x_k}\frac{\partial^2{\cal L}} 
{\partial \phi^\alpha_j\partial\phi^\beta_k}=0\label{multi} 
\end{equation} 
and the equations which determine the universal solution take the form 
\begin{equation} 
\sum_ix_iF^\alpha_i(\phi^\beta)= {c^\alpha}.\label{form} 
\end{equation} 
Here the arbitrary functions 
$F^\alpha_i$ may be regarded as matrix valued functions of all 
the fields $\phi^\beta$ and the constants $c^\alpha$ 
depend upon $\alpha$. These equations, linear in $x_i$ 
provide an implicit solution for the unknowns $\phi^\beta$ 
The proof is along similar lines to the previous case. 
Differentiating (\ref{form}) twice with respect to $(x_j,\;  x_k)$ gives 
\[ 
\left(\sum_i\frac{\partial^2F^\alpha_i}{\partial\phi^\sigma\partial\phi^\tau} 
x_i\right)\phi^\sigma_j\phi^\tau_k+\frac{\partial F^\alpha_j} 
{\partial\phi^\sigma}\phi^\sigma_k+\frac{\partial F^\alpha_k}{\partial\phi^\sigma}\phi^\sigma_j 
=-\left(\sum_i\frac{\partial F^\alpha_i} 
{\partial\phi^\sigma}x_i\right)\phi^\sigma_{jk}. 
\] 
This implies that $\phi^\beta_{jk}$ has the structure 
\begin{equation} 
\phi^\beta_{jk}= \phi^\beta_jG_k+\phi^\beta_kG_j,\label{structure} 
\end{equation} 
where $G_j$ are functions of $\phi^\sigma$ 
and their derivatives whose precise form is unnecessary for the proof. 
 Differentiating (\ref{hom}) yields 
\begin{equation} 
\sum_j\phi^\alpha_j\frac{\partial^2 {\cal L}}{\partial\phi^\beta_j\partial\phi^\gamma_k}+ 
\delta^{\alpha\gamma}\frac{\partial {\cal L}}{\partial\phi^\beta_k}= 
\delta^{\alpha\beta}\frac{\partial {\cal L}}{\partial\phi^\gamma_k}. 
\label{second} 
\end{equation} 
When the result for the form of $\phi^\beta_{jk}$ 
is substituted into the equations of motion (\ref{multi}) the 
consequences of homogeneity (\ref{second}) 
then imply that the equations of motion are satisfied identically.

\section{Iterated Lagrangians} 
In this section it is demonstrated that the universal solution (\ref{one}) is not only a solution to 
the equation of motion derived from any Lagrangian of weight one, but also to that arising from 
iterations of this Lagrangian. If ${\cal E}$ denotes the Euler operator 
\begin{equation} 
{\cal E}=-\frac{\partial}{\partial\phi} 
 +\partial_i \frac{\partial}{\partial\phi_i}-\partial_i\partial_j 
 \frac{\partial}{\partial\phi_{ij}}+\cdots. 
\label{elop} 
\end{equation} 
(In principle the expansion continues indefinitely  but it is sufficient for 
our purposes to terminate at the stage of second derivatives  $\phi_{ij}$, 
since it turns out that the iterations do not introduce any derivatives 
higher than the second thanks to the weight one requirement on ${\cal L}$.) 
 
Then the $r$ fold iteration, defined by 
\begin{equation} 
 {\cal L}^{r}={ \cal L} {\cal E}{\cal L}{\cal E}{\cal L }\cdots 
{\cal L}{\cal E}{\cal L},\label{iter} 
\end{equation} 
where ${\cal E}$ acts on everything to the right gives rise to the generic equation of motion 
\begin{equation} 
\epsilon_{i_1i_2\dots i_{r+1}}\epsilon_{j_1j_2\dots j_{r+1}} 
\frac{\partial^2{\cal L}} 
{\partial \phi_{i_1}\partial\phi_{j_1}}\frac{\partial^2{\cal L}} 
{\partial \phi_{i_2}\partial\phi_{j_2}}\cdots 
\frac{\partial^2{\cal L}}{\partial \phi_{i_{r+1}}\partial\phi_{j_{r+1}}}\det\left\vert 
\frac{\partial^2{\phi}}{\partial x_{i_\alpha}\partial x_{j_\beta}}\right\vert=0,\label{iter2} 
\end{equation} 
where a summation over all choices of $r+1$ out of the $n$ 
variables $x_{i_1},x_{i_2},\dots x_{i_n}$, and similarly $r+1$ out 
of the $n$ variables $x_{j_1},x_{j_2},\dots x_{j_n}$ is implied, 
and the determinant is that of the corresponding $(r+1)\times 
(r+1)$ matrix. Then using the structure of $\phi_{ij}$ 
(\ref{result}) implied by the universal solution, together with 
the homogeneity property it is straightforward to prove that the 
universal solution also solves each member of (\ref{iter2}). In 
fact after summation of each term in the determinantal expansion 
of (\ref{iter2}) over all permutations of the indices, it is 
readily seen that every such sum vanishes. The culminating 
equation, corresponding to the $(n-1)$st equation of motion is the 
Universal Field Equation (so-called because it is independent of 
the choice of initial Lagrangian) which was introduced 
in~\cite{gov}. In this paper it was already noted that the 
universal solution provides a class of solutions 
 of this equation. Moreover, 
in the iteration (\ref{iter}) each successive factor ${\cal L}$ 
may be replaced by a~different function of weight one in $\phi_j$, 
viz 
\begin{equation} 
 {\cal L}^{r}={ \cal L}_{r+1} {\cal E}{\cal L}_r{\cal E}{\cal L}_{r-1} \cdots 
{\cal L}_2{\cal E}{\cal L}_1\label{iter3} 
\end{equation} 
 with the same result: the universal solution is a solution to the resulting equation of motion. 
 
\section{Implications of Lorentz invariance} 
If there is any application of the results of this study, it is necessary to restrict the class of 
Lagrangians further. If one asks for the obvious requirement that the initial 
Lagrangian should be Lorentz 
invariant in addition to being homogeneous of weight one in field derivatives, 
then the answer is unique, 
up to field redefinitions and is just 
\[ 
{\cal L}=\sqrt{\sum \eta^{\mu\nu}\phi_\mu \phi_\nu}, 
\] 
or 
\[ 
{\cal L}=\sqrt{\sum J^{\mu_1\mu_2\dots\mu_m}J_{\mu_1\mu_2\dots\mu_m}}\] 
in the case of multiple fields. $J_{\mu_1\mu_2\dots\mu_m}$ is a typical Jacobian 
of the fields with respect to the base co-ordinates, and the sum is over 
all combinations of the indices ${\mu_1,\mu_2,\dots,\mu_m}$. 
 These Lagrangians are just the Companion Lagrangians proposed in \cite{baker,linda2} 
as covariant analogues of the Klein Gordon Lagrangian and its extension to several fields. 
 
\section{The secret revealed!} 
In order to obtain more insight into the nature of the universal solution, 
it is illuminating to transform to a 
first order formulation. Suppose one sets  $u_j= \frac{\phi_j}{\phi_n}$, $j\neq n$; then in 
the $\phi$ independent case 
the Lagrangian ${\cal L}(\phi_k)$ can be expressed as 
\[ 
{\cal L}(\phi_k)= \phi_n{\cal K}(u_j), 
\] 
where the $u_j$ satisfy the constraints 
\begin{equation} 
u_j\frac{\partial{u_k}}{\partial x_n}-u_k\frac{\partial{u_j}}{\partial x_n} 
=\frac{\partial{u_k}}{\partial x_j}-\frac{\partial{u_k}}{\partial x_j}\qquad 
 \forall \; j,k =1,\dots, n-1.\label{antisym}\end{equation} 
The equation of motion can be written as 
\begin{gather*} 
\frac{\partial }{\partial x_n}\left(\phi_n\frac{\partial{\cal K}}{\partial \phi_n}+{\cal K}\right)+ 
\sum_{j=1}^{n-1}\frac{\partial }{\partial x_j}\left(\phi_n\frac{\partial{\cal K}}{\partial \phi_j}\right)\\ 
\qquad{}=\sum_{j=1}^{n-1}\left(\frac{\partial{\cal K}}{\partial u_j}\frac{\partial{u_j}}{\partial x_n} 
-\frac{\partial}{\partial x_n}\left(\frac{\partial{\cal K}}{\partial u_j}u_j\right)\right)+\sum_{j=1}^{n-1} 
\frac{\partial}{\partial x_j}\left(\frac{\partial{\cal K}}{\partial u_j}\right)\\ 
\qquad= 
\sum_{j=1}^{n-1}\sum_{k=1}^{n-1}\frac{\partial^2{\cal K}}{\partial u_j\partial u_k} 
\left(u_j\frac{\partial u_k}{\partial x_n}- 
\frac{\partial{u_k}}{\partial x_j}\right)=0. 
\end{gather*} 
If there is to be a universal solution, then since ${\cal K}(u_j)$ is 
completely arbitrary, we must have 
\begin{equation} 
u_j\frac{\partial{u_k}}{\partial x_n}+u_k\frac{\partial{u_j}}{\partial x_n} 
-\frac{\partial{u_k}}{\partial x_j} 
-\frac{\partial{u_j}}{\partial x_k}=0\qquad \forall \; j,k =1,\dots, n-1,\label{sym} 
\end{equation} 
or, in terms of $\phi$, 
\[ 
\phi_{nn}\phi_j\phi_k-\phi_{nj}\phi_n\phi_k-\phi_{nk}\phi_n\phi_j+\phi_{jk}\phi_n^2=0. 
\] 
This partial differential equation is satisfied for all viable choices of the indices 
$(j,k)$ in virtue of (\ref{result}) for an implicit solution of~(\ref{one}). 
Combining (\ref{antisym}) with (\ref{sym}) there results the simpler set of equations 
\begin{equation} 
u_j\frac{\partial{u_k}}{\partial x_n}-\frac{\partial{u_k}}{\partial x_j} 
=0\qquad \forall \; j,k =1,\dots, n-1,\label{entire} 
\end{equation} 
which may be made the basis for the deduction of  the solution given. 
\subsection*{Acknowledgements} 
The author is indebted to the Leverhulme 
Trust for the award of an Emeritus Fellowship and to the Clay 
Mathematics Institute for employment when this work was first initiated. 
 
\label{fairlie-lastpage} 
\end{document}